\documentclass[]{elsarticle}

\usepackage{lineno,hyperref}
\usepackage{amsmath}
\modulolinenumbers[5]
\usepackage{subfigure}
\usepackage{xcolor}
\usepackage[margin=1in]{geometry}

\journal{Journal}

\graphicspath{{figs/}}

\begin{document}

\begin{frontmatter}

\title{A review on the protocols for the synthesis of proteinoids}

\author{Saksham Sharma \footnote{mail to: ss2531@cam.ac.uk; saksham096@gmail.com}, Panagiotis Mougoyannis, Giuseppe Tarabella, Andrew Adamatzky}

\address{Unconventional Computing Laboratory, UWE, Bristol, UK}

\begin{abstract}
Protocells are a type of synthetic cells, which, if engineered to have properties similar to a natural cell, can have immense applications in synthetic biology and bioengineering communities. Proteinoids are one of the leading contenders for protocells discovered by Sidney H. Fox in 1950s as the  protein-like molecules which are made out of amino acids. Proteinoids, if made in a right way, can show electrical excitability patterns on its surface with inflow and outflow of acidic and basic ions giving rise to a oscillatory charge behaviour similar to a neuronal spiking potential. The right protocol for the preparation of proteinoids can be hard to find, given that the literature for proteinoid is widely distributed across different scientific subdisciplines - origin of life, synthetic cell engineering, application of proteinoid NPs etc. This review attempts to enlist most of the relevant protocols published in the literature, each catering to different application, whether motivated by fundamental sciences or basic sciences perspective. The article also suggests the best set of protocol that could be followed by the readers to synthesise proteinoid powder as a potential experimental system for proto-cognitive, cosmetic, biomedical, or synthetic biology applications. An overarching picture of proteinoid as a potential system to study the chemical evolution during the transition from abiotic to prebiotic life, in the history of Earth, is also presented in the end.
\end{abstract}


\end{frontmatter}


\section{Introduction}
In Adamatzky (2021) \cite{adamatzky2021towards}, it was proposed that proteinoids - copolymers made of amino acids, can be modeled as (analog) computers. 

Inspiration for this proposal can be traced back to Harada and Fox \cite{harada1958thermal} where the authors showed that proteinoids show oscillatory electrical activity (similar to neuronal spiking activity), hence proteinoids could be one of the contenders of abiotic material that has proto-cognitive abilities \cite{vallverdu2018slime}. 

In this article, we try to enlist the protocols that are published in the past on preparation of proteinoids. While some protocols are solely on the preparation of proteinoid powder, there are some protocols available for encapsulating proteinoid nanoparticles with additive molecules suitable for particular application (such as, targeting cancerous tumours, abosrbing UV light etc.). We start the article in Section \ref{sec:when} by setting up the context and background of why proteinoids were discovered at the first place - and the answer lies in creating synthetic cells (or protocells) which might have predecessors to first living cells that came into existence billion years ago on Earth. Section \ref{sec:method} enlist 12 different protocols for proteinoid preparation some of which are application-driven and others are purely for the sake of proteinoid and its membrane characterisation (themed around origin of life research). Section \ref{sec:best} attempts to briefly summarise the best out of the protocols discussed in Section \ref{sec:method} in order to make proteinoids in a way that is cheaper, easily accessible, reproducible, and multifarious in applications. Section \ref{sec:how} is on the potential complexity route of molecules or the chemical evolution that happened starting from mineral water to first formation of protocells (proteinoids, in lab) to traverse the trajectory from abiotic to prebiotic and then - life as we see it.

\section{When for the first time life emerged from water?}
\label{sec:when}
In a recent news \& views article,  biologist N. Amy Yewdall commented that ``.. a truly living system has yet to emerge from the test tube" \cite{yewdall2022life}. The contention here is the inherent difficulty that lies in the pursuit of engineering the cells, the fundamental building blocks of life, in a medium which is primarily composed of inorganic matter.  Engineering such synthetic cells, on a bright side, is a global challenge of the 21st century considered insurmountable by synthetic biology and bioengineering communities \cite{comp}. If solved, it could lead to  creation of life-like systems for medical and industrial applications. \newline    

Liquid water allows hydrocarbons and CHONS containing molecules to remain stable for a long period of time. If a large macromolecule consists of both types of these molecules, at a large distance, micellesor liposomes are formed. However, at a small distance, alternating sequences in the form of $\alpha$ helix or $\beta$ sheets are formed. These formations are thermally stable and resistant to any kind of degradation \cite{brack1993liquid}.  \newline

To show the emergence and formation of cells from simple organic molecules, Kurihara \textit{et al.} \cite{kurihara2011self} constructed a model protocell using fatty acids and nucleotides by first heating the solution to 95$^{\circ}$C and then reducing it to 65$^{\circ}$C. Transition from simple organic molecules to proteins occur at high temperatures because of condensation-dehydration of amino acids into peptide blocks with water removed in the process. The removal of water increases the reaction rate. Such conditions are usually present in geothermal water resources, hence, it has been argued to be the primary location where complex organic molecules might have been synthesised. Amino acids have been detected in the solutions of formaldehyde with hydroxylamine, formaldehyde with hydrazine in water solutions with HCN. Harada and Fox \cite{harada1958thermal} performed some pioneering model experiments to polymermise reaction products into peptide chains as the first step towards the inorganic synthesis of the protein. Nakashima \cite{nakashima1987metabolism} showed that when amino acid mixtures were heated to high temperatures up to 170$^{\circ}$C, short protein-like molecules, also called `thermal proteinoids' were formed. \newline

Below is the list of various protocols that have been formulated by researchers all around the world after they were first introduced by Harada \& Fox.

\section{Protocols of proteinoid synthesis}
\label{sec:method}
\subsection{Harada \& Fox, 1957 \cite{harada1958thermal}} 
Heating single amino acids often yield decarboxylation products, diketipiperazines, and tars \cite{vegotsky1958characterization}. However, for the purpose of mimicking the formation of protein-like molecules which might have been formed in the hot water resources, as discussed in the above section, amino acids should rather be copolymerised by heat. Harada \& Fox performed copolymerisation of glutamic acid with other acids and their derivatives. The `other' amino acid could be from the following list:
\begin{enumerate}
    \item Glycine
    \item Diketopiperazine 
    \item Aspartic acid
    \item Asparagine
    \item Alanine
    \item Valine
    \item Leucine
    \item Serine
    \item Phenylalanine
    \item Leucine 
    \item Proline
    \item Lysine mononydrochloride
\end{enumerate}

The procedure involves heating a L or DL glutamic acid with the other amino acid from the list above in nearly equal amounts, once when the amino acids are available in a powdered form. The heating is continued for 30-120 minutes at temperature $160-190^{\circ}$C. The final product is allowed to cool which usually take few minutes. After that, the cooled product is treated with 5 or 10 mL of water. \newline

In more precise sense, the heating is undertaken by preparing 0.01 mole of DL-glutamic acid and mixed or grounded with 0.02 mole of glycine in a pestle and mortar. In a test tube which is open to the atmosphere, the mixture is heated at $175-180^{\circ}$C. During heating, as the mixture melts, it evolves a gas which turns the red litmus paper blue. Heating is stopped after the gas stops getting produced, which happens usually after 50 minutes. Upon turning off the heat, the brown liquid solidifies as it turns colder than before. Upon adding 10 mL of water, the solid changes it colour from white to gray. After being left undisturbed overnight, the solid is placed in a centrifuge and then washed with 10 mL of water and then ethanol. The final polymer obtained in a dried form weighs 0.29 g and yields an intense biuret reaction. The most important step now is the dialysis of the solid for 5 days in Visking cellophane tubing with agitation of bath using magnetic stirrer, and then drying off the solid in a vacuum desiccator. It yields a gelatin-like film weighing 0.09 g. 

\subsection{Fox \& Waehneldt 1968 \cite{fox1968therml}}
It is also possible to produce proteinoids from more than two amino acids. In fact, Fox \& Waeheldt produced proteinoids by combining 11 amino acids in different proportions. To give an example, the authors produced Proteinoid No. 50 from 30 mole \% of L-aspartic acid, 30 mole \% of L-glutamic acid, 4.45 mole \% each of L-alanine, L-arginine, HCl, glycine, L-histidine, HCl hydrate, L-isoleucine, L-leucine, L-lysine, HCl, L-proline and L-valine. \newline

The protocol followed to make the proteinoid is by taking 28-33 g of the mixture described above and heating inside the stream of nitrogen in a flask which is under heat at $192-194 ^{\circ}$ for 6 hours. High temperatures is necessary to ensure that the condensation rate is high and acidic composition is yielded. Upon cooling, the reaction mixture is gently heated with 5-6 parts of water. After this, the solution is dialysed against water in the cold for 3 days, filtered, and then, lyophilised. The final product is washed with water and dried over $P_{2}O_{5}$. One of the result noted in this study is that the acidic proteinoids are insoluble in nature and form particulate units. In fact, the proteinoids which are neutral in composition yield microspheres as opposed to particulates.

\subsection{Ishima, Przybylski \& Fox 1981 \cite{ishima1981electrical}}
In an effort to replicate the formation of membrane with electrical excitability in an artificial cell, Ishima, Przybylski \& Fox used proteinoids and lecithin in a solution of potassium acid phosphate with glycerol. \newline

The protocol involves heating polyamino acids in an oil bath at $190^{\circ}$ for 6 hours. The mixture consists of 50 g of a particular amino acid in combination with 50 g of equimolar mixture of 18 amino acids. The heating is performed under a nitrogen blanket. The product is stirred for 2 hours in water and then filtered. The dissolved fraction is then dialysed for 2 days in cold ambience and the buffer (water) is recycled twice a day. The non-diffusates are lyophilised (freeze dried). \newline

60 mg of proteinoid and 60 mg of lecithin is suspended in 3 mL of potassium phosphate solution with glycerol solution in water (50:50 v/v) that is being heated to boiling. Solution is then allowed to cool to room temperature and allowed to stand for up to 2 days. Upon incubation at room temperature, crusts were precipitated from the supersaturated solution and removed using a pipet. The external solution is removed and replaced with mannitol so that in the case of a leucine-rich equimolar proteinoid, a transparent, invisible, spherical membrane is formed from the surface of the crusts. The final spherules formed are of $60-150$ \textmu m in diameter and can maintain their shape provided that the concentration of mannitol used above is of concentration below 300 mM, and ideally between 150-200 mM.

\subsection{Przybylski \& Fox (1984) \cite{przybylski1984excitable}}
The proteinoids with double membranes and various internal compositions can be termed as `spherules' which can microenapsulate dyes, oxidant-reductant compounds or acceptor-donor substances, and then pack them together. These spherules display electrical polarisation and electrical discharges. \newline

The protocol for the formation of such spherules consists of first heating 2:2:1 proteinoid in an oil bath at 190 $^{\circ}$C for 6 hours. The proteinoid mixture consists of 50 g of an equimolar mixture of 18 amino acids under a nitrogen blanket, stirred mechanically in water for 2 hours and filtered subsequently. Following procedure of dialysis for 2 days in a cold temperature and lyophilisation of the nondiffusate is the same as discussed in Section 3.3. Most importanty, the spherules are prepared by solution of 60 mg of the proteinoid and 60 mg of lecithin in  3 mL of water with glycerol in 1:1 volumetric ratio. Upon the addition of K$_{2}$HPO${_4}$, pH is adjusted to reach 5.8. The solution was boiled in a water bath and heated for 1-2 seconds in an open flame. The tube is then allowed to cool to room temperature. The final product is the vesicales/spherules which have certain electrical properties that can be measured using a high input impedance amplifier and recorded on the oscilloscope.

\subsection{Matsuno (1984) \cite{matsuno1984electrical}}
While the experimental protocols detailed above, if followed properly, can yield the electrical membrane, the excitability characteristics of the membrane and the origins of this is explained quite well by Matsuno \cite{matsuno1984electrical}. \newline

An aqeuous solution of proteinoid can yield phase-separated microspheres. These microspheres adjust their structure and functions electrostatically when the constituent proteinoids are in the ionised form. In a given proteinoid microsphere, the acidic and basic proteinoids have different mobilities because of unequal masses and charges. The basic and acidic proteinoids can be modeled as having certain kind of chemical interaction and continuity of ionic material flow which gives rise to electrical excitability of a form similar to spiking trains. This interaction is the result of inflow and outflow rate of basic proteinoids inside microspheres and the concentration of acidic and basic proteinoids inside microspheres. When the outflow rate of basic proteinoid is higher, after certain time, the acidic proteinoids inside attract the outgoing basic proteinoids to not let them outflow indefinitely, as a result, reversing the flow direction. When the inflow of basic proteinoids is higher, the attractive force does not grow indefinitely, as the amount of acidic proteinoid inside do not keep up with the ever increasing positive charge exhibited by basic proteinoids. Thus, the flow reverses. This oscillation of ionic flow is what is seen as electricla excitability and a fundamental signature of life.

\subsection{Przybylski 1985 \cite{przybylski1985excitable}}
Another important problem of relevance in biology is that of relating the membrane structure to its function, particularly that of the governing physicochemical processes that generate electrical oscillation of excitable cells. Constructionist modelling studies employed in studying proteinoids is one method to explain the electrical acitivity of cells made of thermal polymers. \newline

The thermal polymers are synthesises using equimolar parts if DL-amino acids mixed in a mortar and placed in a 1-1 flask and heated at 210 $^{\circ}$ C under nitrogen. Liquid ammonia is used to solubilise the polymer in a dry-ice bath. The product is filtered through a 0.22 \textmu m millipore filter and then, lyophilised. After heating at 188$^{\circ}$C, the product p(asp:glu) is made after 6 hours of heating. On the other hand, p(1 asp: 2 glu: 4 lys: 4 leu: 2 pro) is prepared at 195$^{\circ}$C during 24 hours of polymerisation, and then resolubilised and polymerised during 6 hours at 235-240$^{\circ}$C. \newline

After the preparation of thermal polymers, the aqueous solution of polymers made using 20 mg of polymer dissolved in 1 mL of water, alongside addition of 0.2 mL of glycerol and 20 mg of lecithin, or without lecithin. After heating the solution for 20 s, the solution is allowed to cool slowly. In certain cases, KCl is added to a final concentration of 60 mM. \newline

Vegetable lecithin of concentration 50 mg in 1 mL octane is used to make planar membranes, and p(1 asp: 2 glu: 4 lys: 2 leu: 2 pro) in dimethylformamide (DMF) (50 mg per 1 mL) is added in a teflon chamber with orifice 1 mm in diameter for the membrane.

\subsection{Kolitz-Domb \& Margel (2018) \cite{kolitz2018recent}}
\begin{enumerate}
    \item \textbf{Fluoroscent proteinoid nanoparticles for cancer diagnostics:} To detect colorectal cancer at an early stage, fluorescent imaging allows deep penetration into biological matrices with a higher signal relative to the background noise. Near the infrared spectrum, 700-1000 nm, fluorescent materials allow both deeper penetration and low background signal because of low autofluorescence of body ingredients here. Functional nanoparticle-based NIR probes have enhanced biocompatibility, photostability, easier bioconjugation of biomolecules to the functional groups on the nanoparticle surface, and improved fluorescence signal, over free NIR dyes. To build such a probe, poly(L-glutamic acid-L-phenylalanine-PLLA), or P(EF-PLLA) composed of glutamic acid, phenylalanine is chosen as the base material (proteinoid) to ensure rigidity and biodegradibility. \newline
    
    The NPs are conjugated to the bioactive targeting molecules peanet agglutinin (PEA) and anti-cacinoembryonic antigen antibodies (anti-CAR). Both of these non-conjugated and bioactive-conjugated NIR fluorescent P(EF-PLLA) NPs are found to have an ability to detect color cancerous tumours, as demonstrated in tumor implants in a chicken embryo model. A very high fluorescent signal is provided by the bioactive-conjugated P(EF-PLLA) nanoparticles with upregulated receptors. \newline
    
    \item  \textbf{Proteinoid nanoparticles loaded with anticancer doxorubicin:} Anticancer drug, doxorubicin, is encapsulated inside basic proteinoid nanoparticles. The protocol to do that is by selecting four L-amino acids: lysine, arginine, histidine, and phenylalanine, alongside the addition of PLLA. The main monomer in the complex is lysine giving proteinoid a basic nature. Histidine and arginine are combined together to promote endosomal escape alongside increasing the proficiency of penetration through the endocytic pathway. The smallest size of the nanoparticle achieved for poly(L-lysine-L-arginine-L-histidine-L-phenylalanine-PLLA), with the shortform P(KRHF-PLLA), with 2:3 weight ratio of histidine:phenylalanine, is 36.2$\pm$6.9 nm. The proteinoid NPs have small size distribution and are relatively uniform, and are loaded with 15\% Dox, as opposed to commerically available Doxil with 12.5 \% Dox. Entrapment efficacy of Dox within the NPs is 93\%. \newline
    
    \item \textbf{Proteinoids for cosmetic applications:} For cosmetic applications, it is possible to encapsulate all-trans retinoic acid (at-RA), which is an active form of vitamin A used in cosmetics, within the proteinoid NPs. \newline
    
    The protocol involves synthesising proteinoids from the following amino acids: glutamic acid; phenylalanine; tyrosine, alongside an additional non-alpha amino acid, para-aminobenzoic acid (PABA). Glutamic acid and phenylalanine are selected to grant acidic nature to the final product alongside providing mechanical strength, and an addition of tyrosine for strength enhancement. On its own, at-RA is light-sensitive and degrades to less potent forms in direct contact with the light, however, designing proteinoid NPs with UV absorbing property can protect at-RA from the possibility of degradation. By adding PABA to the proteinoid polymeric chain, alongside tyrosine, yields a proteinoid with wider UV absorption capability (up to 300 nm). \newline
    
    \item \textbf{Proteinoids for anti-fog applications:} For anti-fog applications, the materials often required are the ones with special amphiphilic character. One of the often used anti-fog molecule is surbitan monooleate (SMO). However, it is possible to use proteinoid itself as an anti-fog material with low water contact angle and low roughness. \newline
    
    The proteinoid for the anti-fog application is composed of glutamic acid, phenylalanine, and isoleucine. These proteinoids, poly(L-glutamic acid-L-isoleucine), poly(L-glutamic acid-L-isoleucine-L-phenylalanine), poly(L-glutamic acid-L-phenylalanine), P(EI), P(EIF), P(EF), are self-assembled into NPs, with or without encapsulated SMO molecule. The smalles particles are of the size 89.7$\pm$6.0 nm for the hollow NPOs and 166.1$\pm$2.0 nm for the SMO-filled NPs. \cite{kwon2018disulfide}
    
\end{enumerate}

 \subsection{Kwon, Park \& Kim (2018) \cite{kwon2018disulfide}}
 Proteinoids could also act a carrier of a certain kind of drug and be programmed to release it depending on the external conditions. It is possible by introducing disulfide bond in the proteinoid that lets it assemble in an aqueous phase. In an external environment which is reducing in nature, the disulfide bond would be broken and the micelle is likely to be loosened. As a result, the payload (drug) inside the micelle would be released.  \newline
 
 To prepare proteinoid for the application highlighted above, Asp and Leu (Prot(Asp-Leu)) and Asp, Leu, and DTPA (Prot(Asp-Leu-DTPA)) is prepared by melt-condensation method. 16 mL of glycerol is added to the mixture of Asp (20 g), Leu (4 g), and DTPA (0 or 10 g) and mixed using a spatula inside a flask. Nitrogen is purged inside the flask, and then the flask is closed with rubber plugs, and soaked in an oil bath for 9 h at 170$^{\circ}$C with reflux. Then the reaction mixture is cooled at a room temperature and dissolved in 50 mL of sodium bicarbonate solution (10\%(w/v)). The reaction mixture is dialysed in 3 L of distilled water using a dialysis bag of MWCO 1000 for 4 days. The dialysis medium is changed 7 times for the effective and fast purification and finally, the proteinoid powder is obtained by freeze-drying the dialysed solution. \newline
 
 50 mg of proteinoid prepared above is dissolved in 5mL distilled water. Also, 50 mg of DOX and amaranth are dissolved in 5 mL of DMF. The latter solution is mixed inside the proteinoid solution using magnetic bar and the solution is stirred overnight. Dialysis is performed using a bag of MWCO 3,500 to remove organic solvent and the free solutes. Amount of DOX and amaranth loaded inside the proteinoid micelles is determined using an aliquot amount of dried micelles which are put inside 2 mL of DMF contained in a 5 mL vial, stirred using magnetic bar for 3 h at room temperature, filtered using syringe filter (220 nm), and filtrates assayed spectrometrically. Fluoroscence intensity at 595 nm and 520 nm of the filtrate  is used to determine the amount of DOX and amaranth respectively on a fluoroscence spectrophotometer. The weight of payloads in the micelles is reported as specific loading, defined by the percent of the mass of payload loaded in micelle based on the mass of micelle. 
 
\subsection{Lugasi, Grinberg \& Margel (2020) \cite{lugasi2020designed}}
Cannabidiol (CBD) has many therapeutic and antitumour properties as it can produce anti-proliferative, pro-apoptotic, cytotoxic, anti-invasive, anti-angiogenic, immunomodulatory effects by activating the signaling pathways at different concentrations and generating chemo-preventive properties in various cancer types (breast, colon, lung, prostate, skin, brain, leukaemia, cervical carcinoma). \newline

While the synthesis of Poly(RGD) (P(RGD)) proteinoid is carried out using a mixture of 5 g of the amino acids D-Arg, Gly, and L-Asp with a weight ration of 1:1:1 heated at 180 $^{\circ}$C. The mixture is stirred at 150 rpm for 30 min and allowed to cool at room temperature. The product is cooled and turned into a highly viscous amber-brownish past. The residue is extracted using 30 mL distilled water and freeze-dried to yield a proteinoid polymer powder. The molecular weight and polydispersity index of the proteinoid powder is determined by gel permeation chromatography (GPC) at 25$^{\circ}$ that consists of HPLC pump with a refractive index detector and a Rheodyne injection valve with a 20 \textmu L loop. Proteinoid NPs are prepared using a self-assembly mechanism consisting of 100 mg dried proteinoid powder added to 10 mL NaCl 10 \textmu M aqueous solution and heated to 80 $^{\circ}$ C, stirred at 250 rpm for 20 min until the proteinoid is dissolved completely. 

\subsection{Lugasi \textit{et al.} (2020) \cite{lugasi2020designed}}
Risperidone (RSP) is an antipsychotic drug used to treat bipolar disorder and schizophrenia and can be encapsulated inside the proteinoid NPs for targeted drug delivery which cross the blood-brain barrier and improve pharmacokinetics and drug effectiveness. \newline

Firstly, 100 mg of poly (L-glutamic acid-L-phenylalanine-L-histidine-poly (L-lactic acid)) P(EFH-PLLA) powder is prepared using the typical proteinoid powder protocol described in sections above. It is then added to 10 mL of a 10$^{-5}$ M aqueous NaCl solution, heated to 80$^{\circ}$C and stirred at 250 rpm for 30 min. RSP powder of weight 20\% relative to proteinoid is dissolved in DMSO (0.2 mL) and heated at 80$^{\circ}$C. The RSP solution is added to heated P(EFH-PLLA) solution. Obtained homogeneous solution is allowed to cool at room temperature and form P(EFH-PLLA)/RSP NPs. Extensive dialysis using a cellulose dialysis membrane with MWCO 1000 Da is distilled against water to remove the DMSO. A 3 \textmu m glass microfiber membrane syringe filter is used to remove the untrapped RSP inside the aqueous dispersion of proteinoid NPs.
 
\subsection{Itzhaki \textit{et al.} (2021) \cite{itzhaki2021tumor}} 
The arginine-glycine-glutamic acid (RGD) sequence preferentially adheres to the $\alpha v \beta 3$ integrin and is highly expressed on neovascular endothelial cells that support the tumour growth. It is possible for RGD-based proteinoid NCs to trap and encapsulate a synergistic combination of Palbocic (Pal) and Alpelisib (Alp) and use it to cause reduction of tumour cell growth in various types of cancers.  \newline

The synthesis of proteinoid is similar to the protocol described in the sections above. Hollow P(RGD) proteinoid NCs are prepared by self-assembly process. The process consists of adding 100 mg dried proteinoid polymer to a 28 mL glass screw-capped vial with 10 mL 0.01 mM NaCl aqueous solution. The mixture is heated to 80$^{\circ}$ on a hot plate. The solution is stirred at 250 rpm for 30 min and left on the plate to cool at the RT. P(RGD) NCs which contain various weights of Pal and Alp drugs are prepared by firstly dissolving the proteinoid powder in five glass vials described above (100 mg in 0.01 mM NaCl solution). A DMSO stock solution containing Pal and Alp (each 300 \textmu L) is placed on the hot plate along with the NC vials during constant heating stage at 80$^{\circ}$C. After 30 min of heating, while at 80 $^{\circ}$, stock solution is added to the vial, and Tween 80 surfactant is added to form cloudy dispersions. The clear yellowish solutions are left to cool slowly to room temperature and dialysed against distilled water (using 8000 Da MWCO) to remove excess drugs.

\subsection{Lugasi \textit{et al.(2022)}\cite{lugasi2022chirality}}
Proteinoids have chiral properties which can be studied by performing enantioselective adsorption of L-amino acids at various pH values. The absorption is studied using circular dichroism, isothermal titration calorimetry, and $\zeta-$ potential measurements. \newline

Thermal step-growth polymerisation is carried out by mixing 5 g each  of L-lysine and L- phenylalanine and heating at 140$^{\circ}$C under nitrogen atmosphere, and stirring at 150 rpm for 45 min. The residue upon heating is extracted with 30 mL super-purified water, which is then followed by lyophilisation. Finally, the dried proteinoid powder is obtained. P(KF) NPs are prepared by self-assembly process, the protocol of which is described in the sections above. \newline

P(KF) NPs are titrated against L/D amino acids to undergo enantioselective adsorption. VP-ITC microcalorimeter is used to perform measurement by filling the reference cell with dionised water, sample cell with KF NP suspension in dionised water, and syringe with the (D or L) enantiometric amino acid solution. 

\section{Best practices in proteinoid preparation}

\begin{figure}
  \centering
  \caption{Overview of the protocols published in the literature and discussed in this article on the synthesis of proteinoids}
  \includegraphics[width=1.1\textwidth]{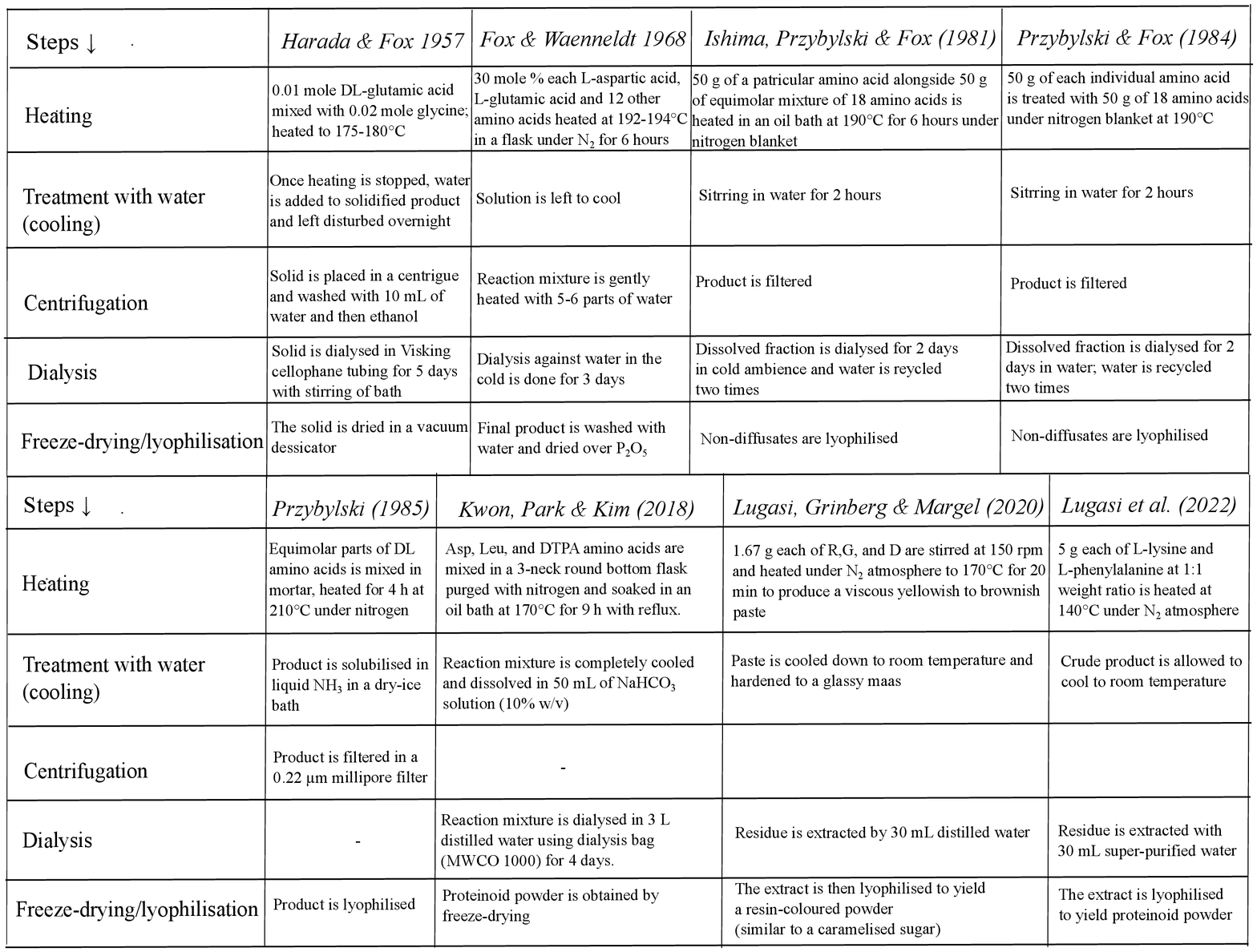}
   \label{tab:tab1}
\end{figure}

\textit{Fox \& Waeheldt (1968)} \cite{fox1968therml}: The protocol for the proteinoid preparation in this article can be seen as quite well studied and detailed in the sense that in each type of proteinoid formed, at least 10 amino acids were used in the mixture. This allows one to vary the amount of amino acids in 10 different ways and thus, form proteinoids which are of acidic, neutral, and basic nature. While the general protocol for synthesis of proteinoids is the same as most other methods (highlighted in Fig \ref{tab:tab1}), a striking conclusion in the article is that: insoluble polymer (in the form of particulate units) is made when the polymer is overall anionic in nature. Basic polymers on the other hand are soluble, and only neutral proteinoids are the ones which are responsible for the formation of microspheres that are useful for proto-cognitive applications as discussed in previous sections. \newline
\textit{Przybylski \& Fox (1983)} \cite{przybylski1984excitable}: While the formation of proteinoid microspheres by ensuring the effective neutral nature of its constituents is a great first step towards studying proto-cognitive abilities in these protocells, what's even more challenging is to ensure that the membranes of these cells have certain form of electrical excitability. Electrical oscillations of any form are usually considered as signatures of intelligence and problem-solving abilities in a biological or proto-biological form, hence, the protocol which can enable that is likely to be considered as an important protocol to note down. Przybylski \& Fox, in their article, mixed proteinoid with lecithin (a mixture of fats) in 3 mL of water with glycerol in 1:1 volumetric ratio. The pH of the solution is increased to 5.8 by adding K$_{2}$HPO$_{4}$ and then boiled in a water bath followed by heating two times for 1-2 s in an open flame. The result is that the proteinoid microspheres have vesicles on their boundary with controlled permeability. Such a membrane-bound proto-cell can easily encapsulate substances such as  $\alpha-$ chlorophyll for weeks, such that the encapsulating capacity of the vesicle is directly correlated to the high membrane potential.

\label{sec:best}

\section{How did life emerge from water?}

It is possible to make elementary life forms in the lab using proteinoid in hot water solutions. It is possible by applying gas electric discharge with coronal spectral analyses. It has been argued that the first living structures were probably formed in hot mineral water with metal ions such as Na, Ca, Mg, Zn, K etc. If the water drops are heated to boiling point on the electrode of the device during gas coronal electrical discharge experiment, it is likely that the conditions are quite similar to how it might be in the primordial atmosphere. Self-organisation of proteinoid microspheres (5-10 \textmu m) could lead to formation of a relatively large organised structure (12-14 mm). Reported in Ignatov \& Mosin \cite{ignatov2013modeling}, these structures once formed, can be preserved and remain unchanged in size for 2.5 years without electrical discharge. Such membrane like structures would exchange energy with the environment and be composed of dissipative structures inside that lead to changes in entropy and formation of stable boundaries over a long period of time.

\label{sec:how}

\section*{}
This project has received funding from ERC Grant No. XXX

\bibliographystyle{elsarticle-num}
\bibliography{proteinoid}

\end{document}